\documentclass[useAMS,usenatbib]{mn2e}
\usepackage{graphicx}
\usepackage{xcolor}

\title[]{Magnetic field tomography, helical magnetic fields and Faraday depolarization}
\author[]{
C. Horellou$^{1}$\thanks{E-mail: cathy.horellou@chalmers.se} 
and 
A. Fletcher$^{2}$\\
$^{1}$ Department of Earth and Space Sciences, Chalmers University of Technology, Onsala Space Observatory, SE-439 92 Onsala, Sweden\\
$^{2}$ School of Mathematics and Statistics, Newcastle University,
        Newcastle-upon-Tyne NE1 7RU, U.K.}
\begin{document}

\date{Accepted 2014 April 6.  Received 2014 March 24; in original form 2014 January 16}

\maketitle

\label{firstpage}

\begin{abstract}
Wide-band radio polarization observations offer the possibility to recover information about the magnetic fields in synchrotron sources, such as details of their three-dimensional configuration, that has previously been inaccessible. 
The key physical process involved is the Faraday rotation of the polarized emission in the source 
(and elsewhere along the wave's propagation path to the observer). 
In order to proceed, reliable methods are required for inverting the signals observed in wavelength space 
into useful data in Faraday space, with robust estimates of their uncertainty. In this paper, we examine how variations of the intrinsic angle of polarized emission $\psi_{0}$ with the Faraday depth $\phi$ within a source affect the observable quantities. Using simple models for the Faraday dispersion $F(\phi)$ and $\psi_{0}(\phi)$, along with the current and planned properties of the main radio interferometers, we demonstrate how degeneracies among the parameters describing the magneto-ionic medium can be minimised by combining observations in different wavebands. 
We also discuss how depolarization by Faraday dispersion due to a random component of the magnetic field attenuates the variations in the spectral energy distribution of the polarization and shifts its peak towards shorter wavelengths. This additional effect reduces the prospect of recovering the characteristics of the magnetic field helicity in magneto-ionic media dominated by the turbulent component of the magnetic field. 
\end{abstract}


\section{Introduction}   

A new generation of radio telescopes will map the polarization of cosmic radio sources over a large range of wavelengths, from a few centimetres to several metres. 
Since the plane of polarization of a linearly polarized wave is rotated by an amount that depends on the magnetic field and free-electron distributions and the wavelength ($\lambda$), the resulting data will probe both the synchrotron-emitting sources and any intervening magneto-ionic medium in unprecedented detail. 
A useful way to characterize the intrinsic properties of magneto-ionic media is
the Faraday dispersion function\footnote{The term `Faraday spectrum' is sometimes used for the Faraday dispersion function, which can be misleading because it is not a true spectrum (function of frequency).}, $F(\phi)$,  which contains information
on the transverse orientation of the magnetic field ($B_\bot$) and 
on the intrinsic polarized emission as a function of Faraday depth, $\phi$. 
The Faraday depth 
is proportional to the integral along the line of sight $z$
 of the product of the density of thermal electrons, $n_e$, and the component of the magnetic field parallel to the line of sight:
\begin{equation}
\phi(z) \propto \int_z^{+\infty} n_e (z') B_{\parallel}(z') dz'\, ,  
\end{equation}
hence, in principle, $F(\phi)$ can be used to obtain both the perpendicular and the parallel components 
of the three-dimensional magnetic field. 
(Our system of coordinates is such that the origin is at the far end of the source and the observer is located at $+\infty$. A magnetic field pointing towards the observer yields a positive Faraday depth.) 

Reconstruction of $F(\phi)$  
is usually done by taking advantage of the Fourier-transform type relationship between the observed polarized emission and the Faraday dispersion function.
The {\it observed} complex polarization $P(\lambda^2)$ can be expressed as
the integral over all Faraday depths of
the {\it intrinsic} complex polarization $F(\phi)$ modulated by the Faraday rotation
\citep{1966MNRAS.133...67B}:  
\begin{equation}
P(\lambda^2) = 
\int_{-\infty}^{+\infty} 
F(\phi) e^{2{\rm i}\phi \lambda^2} 
d\phi,  
\label{eqpol}
\end{equation}
so that $F(\phi)$ can be expressed in a similar way:
\begin{equation}  
F(\phi) = 
\frac{1}{\pi}
\int_{-\infty}^{+\infty} 
P(\lambda^2) 
e^{-2{\rm i}\phi \lambda^2} 
d\lambda^2  \, . \\ 
\label{eqF} 
\end{equation}

$F(\phi)$ is a complex-valued function:
\begin{equation}
F(\phi) = |F(\phi)| e^{2{\rm i}\psi_0(\phi)} \, , 
\end{equation}
where $|F(\phi)|d\phi$ is the fraction of polarized flux that comes from regions of Faraday depth between $\phi$ and $\phi+d\phi$, 
$\psi_0$ is the intrinsic polarization angle (perpendicular to the transverse component of the magnetic field, $\vec{B_\perp}$) and may itself depend on $\phi$.

Equation~(\ref{eqF}) lies at the heart of methods to recover $F(\phi)$ from multi-frequency observations of the complex polarized intensity 
(called Rotation Measure, RM, synthesis; \citealt{2005A&A...441.1217B}). 
The RM synthesis has been used to recover Faraday components of compact sources 
(e.g. \citealt{2010ApJ...714.1170M}) 
and diffuse structures in the Milky Way 
(e.g. \citealt{2009A&A...494..611S}), 
in nearby galaxies 
(e.g. \citealt{2009A&A...503..409H}) 
and in galaxy clusters (e.g. \citealt{2011A&A...525A.104P}). 
Several techniques have been proposed to deal with the limited $\lambda^2$ coverage provided by real telescopes
(RM-CLEAN; \citealt{2009IAUS..259..591H}; 
sparse analysis and compressive sensing; 
\citealt{2011A&A...531A.126L}, \citealt{2012AJ....143...33A}; 
multiple signal classification; \citealt{2013MNRAS.430L..15A}) 
and with the missing negative $\lambda^2$ 
(e.g. using wavelet transforms; \citealt{2010MNRAS.401L..24F}, \citeyear{2011MNRAS.414.2540F}). 
\cite{2012A&A...543A.113B} also used wavelets to analyze the scales of structures in Faraday space and emphasized the need to combine data at high and low frequencies. 
Because of the difficulty of the RM synthesis technique to recover multiple Faraday components, 
it has been suggested to use direct $q(\lambda^2)$ and $u(\lambda^2)$ fitting, where $q$ and $u$ are the $Q$ and $U$ Stokes parameters normalised to the total intensity $I$ 
(\citealt{2011AJ....141..191F}; \citealt{2012MNRAS.421.3300O}). 

\begin{table*}
\caption{Frequency range and sensitivity of the listed instruments.
More details are given in Sect.~\ref{Observables}. \hfill 
}
\begin{tabular}{ccccccc}
Instrument 	& Note 	& Frequency band	& Sensitivity 	& Channel width &Integration time &Reference\\
		&	& [MHz]			& [mJy]		& [MHz]\\
\hline
JVLA 		&S-band	&2000-4000		&0.3		& 2	& 10 min &(1)\\
JVLA 		&L-band	&1000-2000		&0.6		& 1	& 10 min &(1)\\
ASKAP		&	&700--1800		&2.5 		& 1	& 10 min &(2)\\
SKA1		&Survey	&650--1670		&0.3 		& 1	& 10 min &(3)\\
SKA1		&Mid 	&350--3050		&0.1 		& 1	& 10 min &(3)\\
SKA1		&Low 	& 50--350		&0.08 		& 1	& 10 min &(3)\\
GMRT 		&	&580--640		&0.5  	 	& 1	& 1 hr   &(4)\\
GMRT 		&	&305--345		&3.8	 	& 1	& 1 hr   &(5)\\
WSRT		&92 cm	&310--390		&3.9 		& 1	& 1 hr   &(6)\\
LOFAR 		&HBA2	&210--250		&2.6--6.0	& 1	& 1 hr   &(7)\\
LOFAR 		&HBA1	&110--190		&1.6		& 1	& 1 hr   &(7)\\
\hline
\end{tabular}
\vspace{0.2cm}\vfill
(1)~{\tt science.nrao.edu/science/surveys/vlass/VLASkySurveyProspectus\_WP.pdf}\\
(2)~{\tt www.atnf.csiro.au/projects/askap/spec.html}\\
(3)~{\tt 
www.skatelescope.org/wp-content/uploads/2013/03/SKA-TEL-SKO-DD-001-1\_BaselineDesign1.pdf}, page 18.\\
(4)~{\tt www.ncra.tifr.res.in/ncra/gmrt/gmrt-users}\\
(5)Farnes (private communication).\\  
(6)~{\tt  www.astron.nl/~smits/exposure/expCalc.html}\\  
(7)~{\tt www.astron.nl/radio-observatory/astronomers/lofar-imaging-capabilities-sensitivity/sensitivity-lofar-array/sensiti}, table~4, for a 40-station Dutch array. Those numbers are preliminary, especially in the highest band.
\label{tab1}
\end{table*}

In this paper we 
show how observations, performed in the various wavelength ranges available at existing and planned radio telescopes, can be used to constrain the variation of $\psi_{0}$ (and therefore the orientation of the magnetic field component perpendicular to the line of sight) with $\phi$. 
We use a Fisher matrix analysis to quantify the precision that can be achieved for 
fitted parameters and investigate the degeneracies that exist between the different constituents of our model. 
Recently, 
\cite{2014PASJ...66....5I} 
performed a similar analysis to evaluate the capability of new radio telescopes 
to constrain the properties of intergalactic magnetic fields through observations of background polarized sources. 
Their work assumed two Faraday components, each with a constant $\psi_{0}$, 
a narrow one (the compact radio source)
and a broad one (possibly associated with the Milky Way). 
Here we consider {\em a linear variation of $\psi_{0}$ with $\phi$} and 
show how the degeneracies between pairs of model parameters can be broken using complementary datasets from different instruments in order to recover $\psi_{0}(\phi)$, using  
two simple models of $F(\phi)$, a constant  
and a Gaussian. 

In the simple cases we consider, the variation of $\psi_0(\phi)$ can be produced by a helical magnetic field. 
Magnetic helicity is a natural consequence of dynamo action and sophisticated statistical methods 
have been 
devised to try to infer its presence, although without inclusion of Faraday effects
(\citealt{2011A&A...530A..89O}, \citealt{2011A&A...530A..88J}). 
Anomalous depolarization (an increase rather than the usual decrease of the degree of polarization with wavelength) produced by an helical field was discussed by 
\cite{1998MNRAS.299..189S}. 
Helical fields have been invoked to explain the anomalous depolarization properties of the nearby galaxy NGC~6946 
(\citealt{1997A&A...326..465U}) 
and polarization characteristics of the central part of the starburst galaxy NGC~253
(\citealt{2011A&A...535A..79H}). 
Helical magnetic fields are also important in galactic and protostellar jets 
(e.g. \citealt{2009pjc..book..555K}, 
\citealt{2011ApJ...737...43F}). 
Bi-helical fields (with opposite signs of helicity on small and large scales) 
are produced in simulations of galactic dynamos and 
the signatures of such fields 
are discussed in a recent paper 
by \cite{2014arXiv1401.4102B}. 
In this paper, we focus on the detectability of single-helical magnetic fields.

\section{Analysis}

\subsection{Observables} 
\label{Observables}

We consider observations of the Stokes parameters $Q$ and $U$ with the instruments listed in Table~\ref{tab1}. 
We used a nominal integration time of 1 h for the low-frequency observations (Giant Meterwave Radio Telescope, GMRT, 
Westerbork Synthesis Radio Telescope, WSRT, Low Frequency Array, LOFAR) 
and 10 min for observations with the more sensitive instruments (Jansky Very Large Array, JVLA, Australian Square Kilometre Array Pathfinder, ASKAP and Square Kilometre Array 1, SKA1). 
This allows an easy comparison of the sensitivities and makes it possible to display the confidence intervals of 
the parameters of interest on a common graph (Figs~\ref{fig2} and \ref{fig3}). 
We used a channel width of 1~MHz for all instruments except the JVLA 
for which a channel width of 2~MHz is more than sufficient in the wide S-band 
(2000--4000~MHz) to resolve the main features of the spectral energy distribution of the polarization. Note that all instruments allow the use of narrower channels; however, there is an obvious trade-off between sensitivity per channel  
and total integration time. We have varied the channel width over two orders of magnitude between 0.1 and 10~MHz 
and observed that the resulting precision on the main parameter of interest, $\beta$ (equation~(\ref{eqbeta})), changes by less than $10^{-2}$ 
for a same total integration time 
of the SKA1-Survey. 
The quoted sensitivities are indicative 
since several instruments listed in Table~\ref{tab1} are still in their design phase. 
Also, some bands, especially the low-frequency ones, will be affected by radio frequency interferences and a fraction of the channels will be missing. With real data at hand it will be straightforward to include the actual frequency coverage and sensitivities in the modeling of a particular data set. 

We scaled the sensitivities $\sigma_{\rm lit}$ quoted in the literature for a given effective bandwidth $BW_{\rm lit}$ and integration time $t_{\rm lit}$ to new values of the channel width $\Delta\nu$ and integration time $t_{\rm int}$, as given in the table, for a given number of tunings $N_{\rm tunings}$ to cover the whole bandwidth: 
\begin{equation}
\sigma = \sigma_{\rm lit} \sqrt{\frac{BW_{lit}}{\Delta\nu}    
\frac{t_{\rm lit}}{t_{\rm int}} N_{\rm tunings} \, . 
} 
\end{equation}

The JVLA will be used to carry out    
sensitive surveys of large parts of the sky. We use figures provided by 
Steven T. Myers [National Radio Astronomy Observatory (NRAO)] in the Karl Jansky VLA Sky Survey Prospectus (see Table~\ref{tab1}) 
for the JVLA in its B-configuration. 
In the S-band (2--4~GHz), the effective bandwidth 
(free from radio frequency interferences) 
is 1500~MHz, and a noise level of 0.1~mJy can be achieved in 7.7~seconds. 
The size of the synthesized beam is $2.7''$ at the centre of the band. 
In the L-band (1--2 GHz), the effective bandwidth 
is 600~MHz and a noise level of 0.1~mJy can be achieved in 37 seconds of integration. 
The size of the synthesized beam is $5.6''$. 
In both cases we assumed a single tuning. 
We note that \cite{2014arXiv1401.1875M} 
recently submitted a science white paper for a JVLA sky survey in the S-band (in the C-configuration),  
in which they consider several alternatives ranging from a shallow all-sky survey to ultra-deep fields of a few tens of square degrees. They estimate that a shallow all-sky survey of a total of about 3000~hours would lead to the detection of over $2\times10^5$ polarized sources. 

According to the ASKAP website, 
a continuum sensitivity of 29 to 37 $\mu$Jy~beam$^{-1}$ for beams between 10$''$ and 30$''$ can be reached in 1 hour for a bandwidth of 300~MHz. Four tunings would be required to cover the whole frequency band from 700 to 1800~MHz, 
so in a total of 1 h 
a noise level of 1-- 1.3~mJy per 1~MHz channel would be reached. 
A major polarization survey with ASKAP (POSSUM) is in the design study phase. 

In its first phase, the SKA will observe at low frequencies (50 -- 350 MHz, SKA1-Low), mid-frequencies (0.35 -- 3.05~GHz, SKA1-Mid) and in a survey mode in the 0.65 -- 1.67~GHz range (SKA1-Survey). 
In 1 hour of observation and per 0.1~MHz channel, the sensitivity is expected to be 
63~$\mu$Jy for SKA1-Mid, 
103~$\mu$Jy for SKA1-Low, and
263~$\mu$Jy for SKA1-Survey. 
We have assumed a single tuning for SKA1-Low and that for SKA1-Mid four tunings (in a 770~MHz bandwidth each) will be needed to cover the whole band; 
for SKA1-Survey, the maximum bandwidth will be 500~MHz, so two tunings will be needed.
The corresponding noise levels per 1~MHz bandwidth and after 10 minutes of observations are given in Table~\ref{tab1}.   

For the GMRT 610~MHz band, our sensitivity estimate is based on the figures quoted by
\cite{2013arXiv1309.4646F} who reached a noise level in $Q$ and $U$ of
36~$\mu$Jy per beam of 24$''$ in 180 minutes in a 16~MHz band centered at 610~MHz.
Four tunings would be required to cover the whole band. 

Our estimate of the sensitivity of the GMRT in the 325~MHz band relies on a noise level of 2.7~mJy per beam per 1~MHz channel in 1~hour, based on polarization observations of a pulsar done in 2011 
(Farnes, private communication) and assuming that all 30 antennas would be available. 
Assuming that 2 tunings would be necessary to cover the whole band, this gives a noise level of 3.8~mJy in a total of 1~h. 

The WSRT also operates in the 320~MHz band (called the 92~cm band). After 1 hour of observation, the theoretical noise level in Stokes $I$ is 1.2254~mJy~beam$^{-1}$ in a 10~MHz band.
This corresponds to about 3.9~mJy~beam$^{-1}$ in a 1~MHz channel. 
Note that confusion noise is expected to be significant in observations of the Stokes parameter $I$ but it can be neglected $Q$ and $U$.
\cite{2013A&A...559A..27G} 
recently detected polarization with the WSRT towards the Andromeda galaxy at 350~MHz.  

\begin{figure*}
\includegraphics[width=0.47\textwidth]{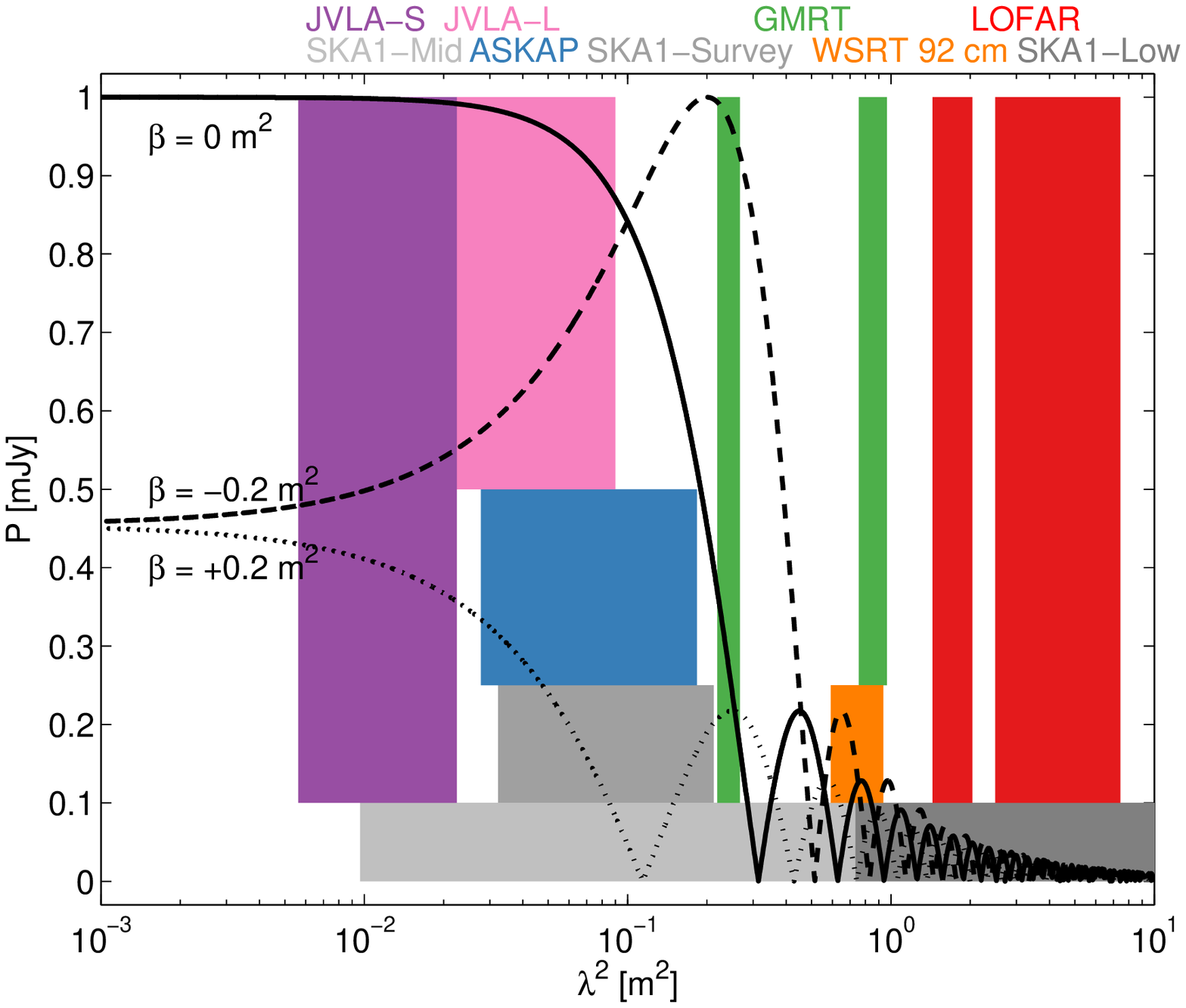}
\includegraphics[width=0.47\textwidth]{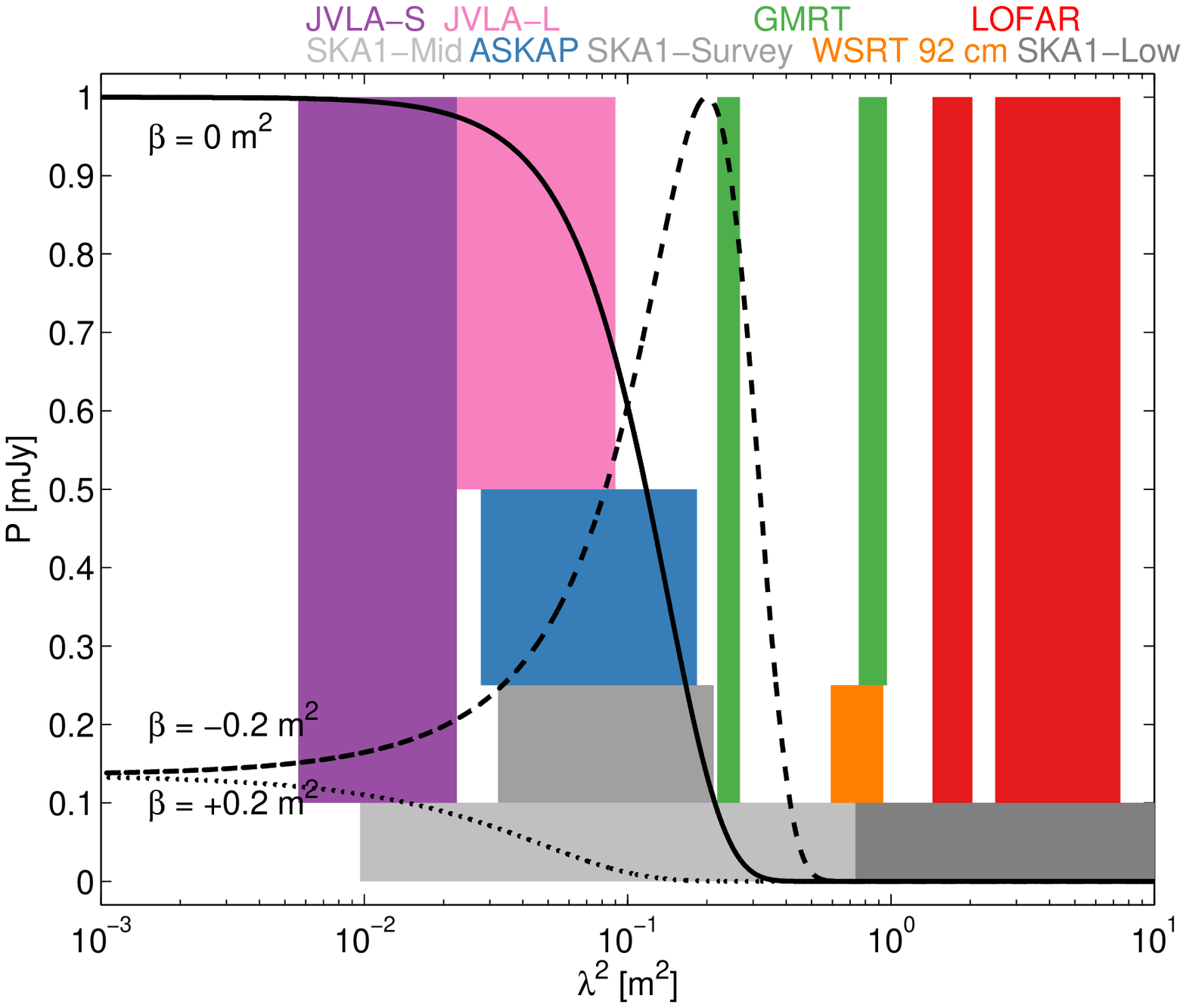}
\includegraphics[width=0.47\textwidth]{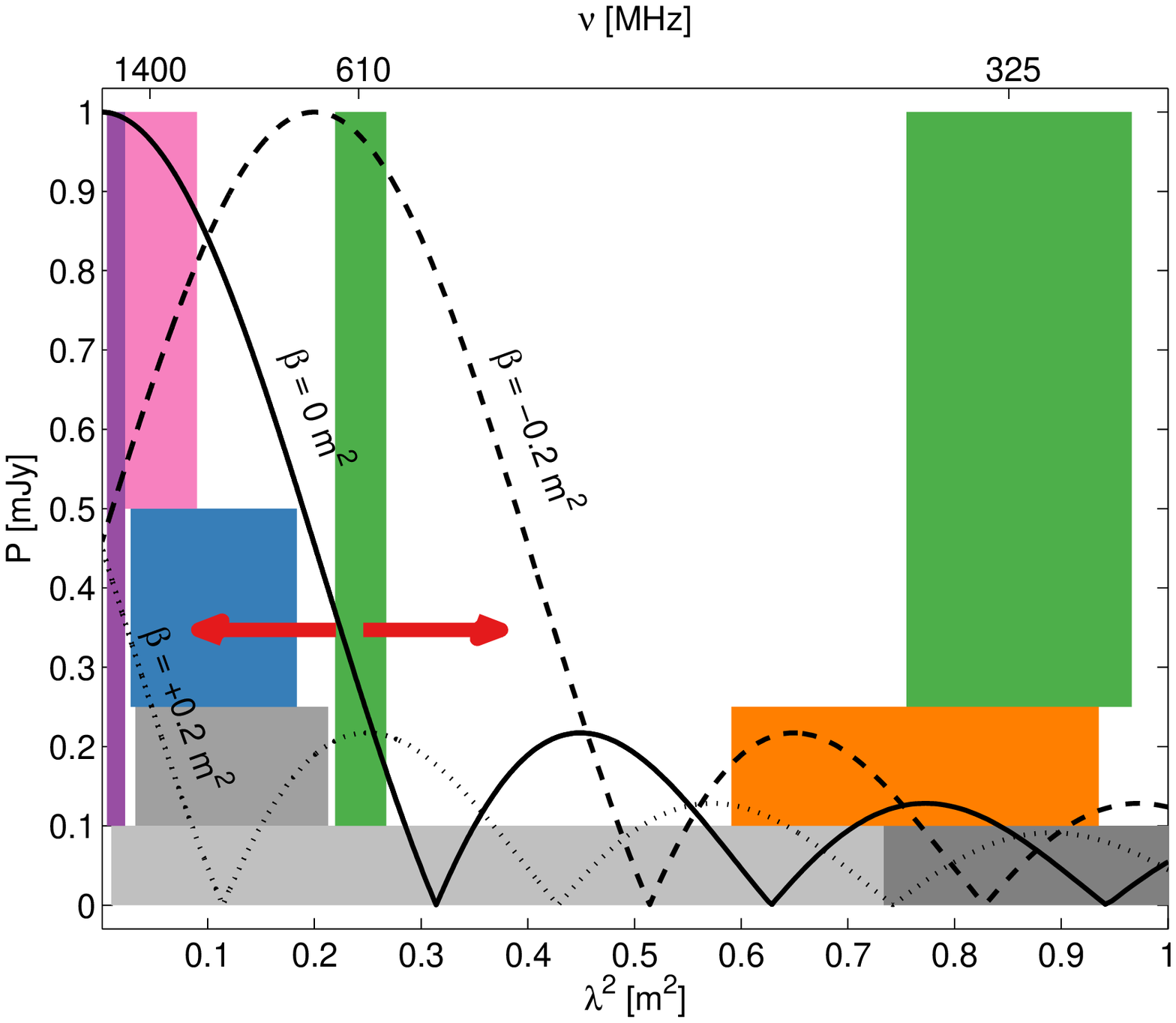}
\includegraphics[width=0.47\textwidth]{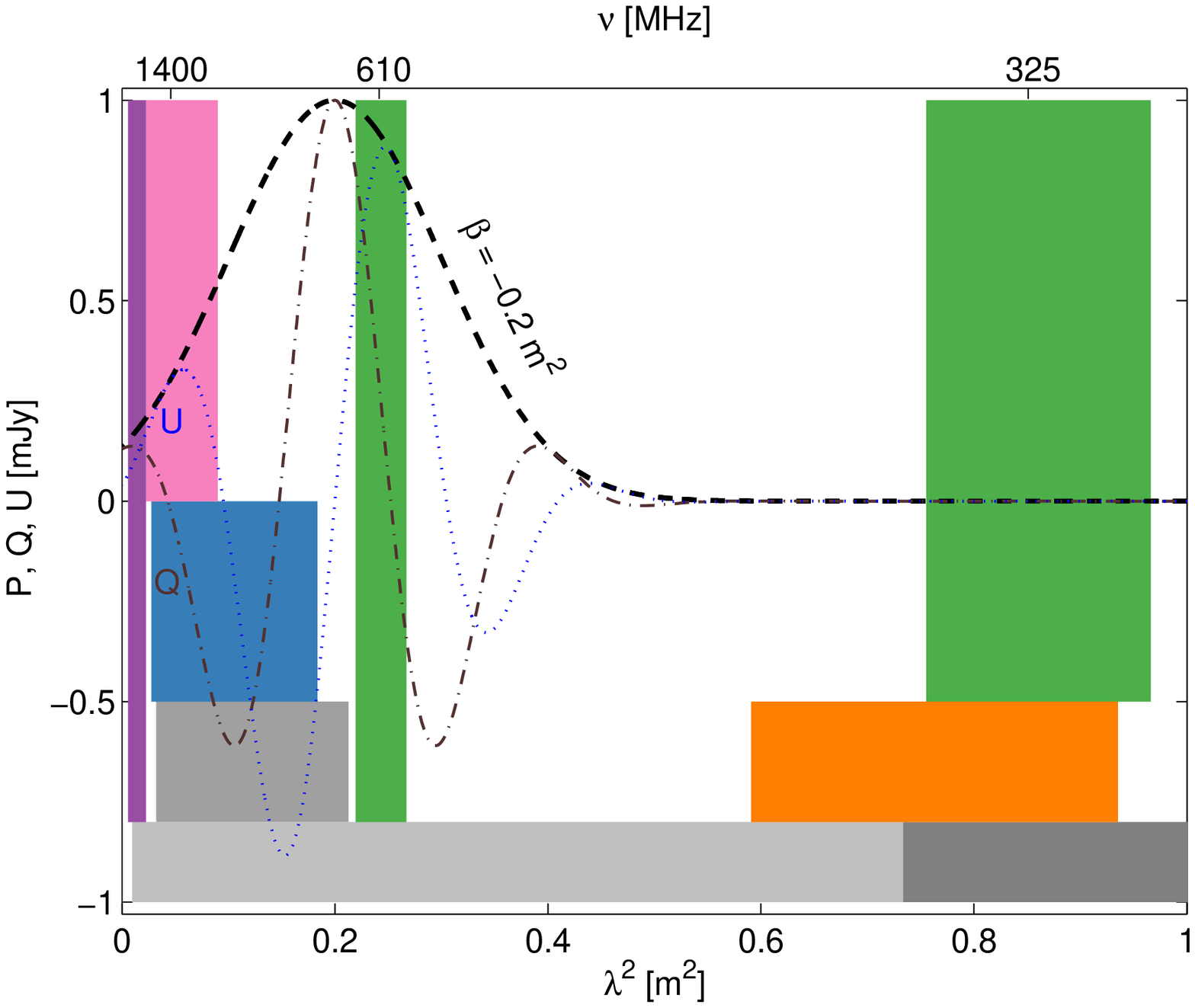}
\caption{Polarized intensity versus $\lambda^2$ for the top-hat model of 
the Faraday dispersion function (left) 
and the Gaussian model (right) 
 for different values of the $\beta$ parameter: 
$\beta =0$ (solid line),
$\beta = -0.2$~m$^2$ (dashed line),
and $\beta = +0.2$~m$^2$ (dotted line). 
The graphs on the top row are in semilogarithmic scale and extend out to the frequency bands of LOFAR, while the graphs on 
the bottom row are in linear scale and extend to a wavelength of about 1 metre, a band covered now by the GMRT 
and the WSRT and to be covered by the SKA. 
The wavebands of several instruments are shown in colour. 
For the Gaussian model, the variations of the Stokes parameters $Q$ and $U$ are also shown (bottom right panel, brown dotted-dashed and blue dotted lines) for the case with 
$\beta = -0.2$~m$^2$. 
The pattern of polarized intensity is shifted horizontally as $\beta$ varies, 
peaking at $\lambda^2 = -\beta$, as illustrated by the red arrows in the bottom left panel. A negative $\beta$ ($B_{||}$ and handedness of $B_\bot$ of opposite signs) results in a peak at $\lambda^2 > 0$, 
whereas a positive $\beta$ results in increased depolarisation due to the combined
depolarization effects of intrinsic helicity and Faraday rotation. 
}
\label{fig1}
\end{figure*}

The high-band array (HBA) of the LOFAR operates at frequencies between 110~MHz 
and 250~MHz with a filter between 190 and 210~MHz. 
LOFAR has detected polarization in the HBA and rotation measures could 
be inferred  
(e.g. in pulsars, \citealt{2013A&A...552A..58S}, 
and in polarized sources in the field of M~51, Mulcahy et al., in prep.). 
However, at the LOFAR frequencies depolarization is extremely strong and 
for the fiducial models presented in this paper the measurements at the quoted sensitivities do not provide improved constraints on the parameters related to the magnetic field. 
The LOFAR frequency coverage is displayed in Fig.~\ref{fig1} but the LOFAR confidence intervals are therefore not shown in the other figures. 

\subsection{The Fisher matrix}  

The Fisher analysis is often used in cosmology 
(e.g. \citealt{2006astro.ph..9591A}, page 94). 
Consider a set of $N$ data points and a model with $P$ parameters, $p_1, ...p_P$. 
The Fisher matrix elements $\mathcal{F}_{jk}$ are proportional to the second partial derivatives with respect to two given parameters of the likelihood function $\mathcal{L}$ that the data set derives from the given model.  
If the measurement errors follow a Gaussian probability distribution, then 
\begin{equation}
\mathcal{F}_{jk} = - \frac{\partial^2 \ln \mathcal{L}}{\partial p_j \partial p_k}  
= \frac{1}{2} \frac{\partial^2 \chi^2}{\partial p_j \partial p_k}\, , 
\end{equation}
where $\chi^2$ is defined in Eq.~(\ref{eqchi2}). 
Denoting $Q_{\rm mod}$ and $U_{\rm mod}$ the values of the Stokes parameters $Q$ and $U$ for the assumed model, estimated at wavelengths $\lambda_i$ and with noise levels $\sigma_i$, we have

\begin{equation}
\begin{array}{ll}
\chi^2 = \sum\limits_{i=1}^N 
& \left(
\frac{Q_i - Q_{\rm mod}(\lambda_i; p_1,...,p_P) }{\sigma_i}
\right)^2\\
& + 
\left(
\frac{U_i - U_{\rm mod}(\lambda_i; p_1,...,p_P) }{\sigma_i}
\right)^2 \, .
\end{array}
\label{eqchi2}
\end{equation}
The Fisher matrix elements can be written as 
\begin{equation}
\begin{array}{ll}
\mathcal{F}_{jk} = \sum\limits_{i=1}^N 
\frac{1}{\sigma_i^2} 
&\left(
\frac{\partial Q_{\rm mod}(\lambda_i^2; p_1,...,p_P)}{\partial p_j} 
\frac{\partial Q_{\rm mod}(\lambda_i^2; p_1,...,p_P)}{\partial p_k} 
\right.\\
& +
\left.
\frac{\partial U_{\rm mod}(\lambda_i^2; p_1,...,p_P)}{\partial p_j} 
\frac{\partial U_{\rm mod}(\lambda_i^2; p_1,...,p_P)}{\partial p_k} 
\right) \, .
\end{array}
\end{equation}
The covariance matrix is the inverse of the Fisher matrix: 

\begin{equation}
\sigma^2_{jk} = (\mathcal{F}^{-1})_{jk}  \, .
\end{equation}

\subsection{Models of $F(\phi)$ with linearly varying intrinsic polarization angle}

We consider two simple models for $F(\phi)$, each with a linearly varying $\psi_{0}$ as a function of Faraday depth: 
\begin{equation}
\psi_0(\phi) = \alpha + \beta\phi, 
\label{eqpsi0}
\end{equation}  
 with some constants $\alpha$ and $\beta$.
This is a parametrisation of $\psi_0$ 
as a first-order polynomial and,
as discussed below, it can also be interpreted as
a helical magnetic field.    

\subsubsection{Constant Faraday dispersion in the source}
\label{sect231}

One of the simplest possible models for $F(\phi)$ is the top-hat,  
\begin{equation}
F(\phi; {\bf{p}}) = F_0 \, T(\phi; {\bf{p}}) e^{2{\rm i} \psi_0(\phi;{\bf p})} \, , 
\end{equation} 
where 
the set of parameters is 
${\bf p} = (F_0, \phi_{\rm 0}, \phi_{\rm m}, \alpha, \beta)$, 
$\psi_0(\phi; {\bf{p}})$ is given by Eq.~(\ref{eqpsi0}) and 
$T(\phi; {\bf{p}})$ is the top-hat function, with $T=1$ in the range $\phi_{\rm 0}-\phi_{\rm m}<\phi<\phi_{\rm 0}+\phi_{\rm m}$ and $T=0$ elsewhere. 
The complex polarization is
\begin{equation}
P(\lambda^2; {\bf p}) = 2 \phi_{\rm m} \, F_0 \, {\rm sinc} [ 
2\phi_{\rm m}(\lambda^2 + \beta) ] 
e^{2{\rm i} \psi(\lambda^2; {\bf{p}})}  \, , 
\label{eqPtophat}
\end{equation}
where
\begin{equation}
\psi(\lambda^2; {\bf{p}}) = \alpha + (\lambda^2+\beta)\phi_{\rm 0} 
\label{psiobs}
\end{equation}
and ${\rm sinc}(x) = \sin(x)/x$. 

In a uniform slab, the Faraday depth varies linearly with the $z$-coordinate 
for $\phi$ between $\phi_{\rm 0} \pm \phi_{\rm m}$:
\begin{equation}
\phi = 0.81 n_e B_{\parallel} z \, , 
\end{equation}
where $n_e$ is in cm$^{-3}$, $B_{\parallel}$ in $\mu$G, $z$ in pc and $\phi$ in rad~m$^{-2}$.

Consider a magnetic field with a constant line-of-sight component, but
with a rotating component in the plane of the sky:
\begin{equation}
B =  
\left(
\begin{array}{c}
B_{\perp}\cos(\alpha + k_H z ) \\
B_{\perp}\sin(\alpha + k_H z)\\ 
B_{\parallel}\\
\end{array}
\right) \, . 
\end{equation}
The intrinsic polarization angle clearly varies with the Faraday depth:
\begin{equation}
\psi_0 
= \alpha + k_H z 
= \alpha + \beta \phi, 
\label{eqbeta}
\end{equation}
where
\begin{equation}
\begin{array}{ll}
\beta &= 
\frac{k_H}{0.81 n_e B_{\parallel}} \\ 
&=  0.086\, {\rm m}^2 
\left(\frac{k_H}{2\pi\, {\rm rad~kpc}^{-1}} \right) 
\left(\frac{0.03\, {\rm cm}^{-3}}{n_e}\right)
\left(\frac{3 \,\mu {\rm G}}{B_{||}}\right) \, .
\end{array}
\end{equation}  
For an helical field with $k_H \simeq 2\pi$~rad kpc$^{-1}$, we have 
$\beta\simeq 0.1$~m$^2$. 

Note that the sign of $\beta$ depends on the relative orientation of the magnetic field component along the line of sight and the handedness of the helix. 
A positive $\beta$ means that $B_{\parallel}$, which produces the Faraday rotation, and the intrinsic rotation of $B_{\perp}$ have the same direction. 
A negative $\beta$ means that the Faraday rotation effectively counteracts the intrinsic rotation of the plane-of-sky magnetic field. 
This effect will be discussed further in Sect.~3. 

\subsubsection{Gaussian Faraday dispersion function}
\label{sect232}

As a simple alternative to the top-hat parametrisation of $F(\phi)$ we also consider a Gaussian form,
\begin{equation}
F(\phi; {\bf p}) = F_0 \, \exp
\left[ -0.5 \left(
\frac{\phi-\phi_{\rm 0}}{\sigma_\phi}
\right)^2\right] 
e^{2{\rm i} \psi_0(\phi; {\bf p})}, 
\end{equation}
where ${\bf p}$  
and $\psi_0 (\phi, {\bf p})$ are defined as before. This gives a complex polarization of 
\begin{equation}
P(\lambda^2; {\bf p}) = 
\sqrt{2\pi} \sigma_\phi \, F_0 \exp
\left[ -0.5 \left(
\frac{\lambda^2 + \beta}{(2\sigma_\phi)^{-1}}
\right)^2\right] 
e^{2{\rm i} \psi(\lambda^2; {\bf p})}, 
\label{eqPgaussian}
\end{equation} 
where $\psi(\lambda^2; {\bf p})$ is given by Eq.~(\ref{psiobs}). 
The modulus of $P(\lambda^2)$ is a Gaussian centered at $\lambda^2 = -\beta$ 
with variance $\sigma^2 = (2\sigma_\phi)^{-2}$. 

\subsubsection{General case}

In the two previous sections we calculated $P(\lambda^2)$ by integrating Eq.~(\ref{eqpol}) analytically. 
Using the properties of the Fourier transforms, we now show why 
any linear variation of $\psi_{0}$ with $\phi$ produces a 
translation 
of the observed polarized intensity 
in the $\lambda^2$ space. 

Using the standard expression for Fourier transform (integral over $t$ from $-\infty$ to $+\infty$ of a function $f(t)$ times $e^{-2\pi j\nu t}$ 
for the direct transform, and times $e^{2\pi j\nu t}$ for its inverse), Eq.~(\ref{eqpol}) can be written as 
\begin{equation}
P(\pi \lambda^2) = {\rm FT}^{-1}\{F(\phi)\}, 
\end{equation}
where ${\rm FT}^{-1}$ is the inverse Fourier transform. 

Using 
$|F(\phi)| = F_c(\phi) * \delta(\phi-\phi_{\rm 0})$ where $F_c(\phi)$ is a real-valued function centered at $\phi=0$,
\begin{equation}
P(\pi\lambda^2) = {\rm FT}^{-1}\{ F_c(\phi) *\delta(\phi - \phi_{\rm 0}) \cdot e^{2{\rm i} (\alpha + \beta\phi)}\} \,. 
\end{equation}
The factor $e^{2{\rm i} \alpha}$ is independent of $\phi$ and can be taken out of the
integral. Multiplication becomes a convolution in the Fourier ($\lambda^2$) space and convolution becomes a multiplication, so 
\begin{equation}
P(\pi\lambda^2)  
= e^{2{\rm i} \alpha} {\rm FT}^{-1} \{ F_c(\phi)\} \cdot {\rm FT}^{-1}\{\delta(\phi - \phi_{\rm 0})\} * {\rm FT}^{-1} \{ e^{2{\rm i} \beta\phi}\} \, .
\end{equation}
Translation in the $\phi$-space gives a rotation in $\lambda^2$-space, 
and the inverse transform of the term involving $\beta\phi$ becomes a delta-function, giving
\begin{equation}
\begin{array}{ll}
P(\pi\lambda^2)  
& = 
e^{2{\rm i} \alpha} 
\left( 
{\rm FT}^{-1}\{F_c(\phi)\} 
\cdot e^{2{\rm i} \pi\phi_{\rm 0}\lambda^2} 
\right) 
* \delta(\lambda^2 + \frac{\beta}{\pi}) \, .
\end{array}
\label{eqGeneralcase}
\end{equation}
We then obtain 
\begin{equation}
\begin{array}{lll}
P(\pi\lambda^2) 
& = &
\left( 
{\rm FT}^{-1}\{F_c(\phi)\, e^{2{\rm i} \alpha}\} * \delta(\lambda^2+\frac{\beta}{\pi}) 
\right) 
e^{2{\rm i} \pi\phi_{\rm 0}(\lambda^2+\frac{\beta}{\pi})}  \\
& = & 
P_{\beta=0} (\pi(\lambda^2 + \frac{\beta}{\pi})) 
\, e^{2{\rm i} \phi_{\rm 0}(\pi\lambda^2 + \beta)} \, ,
\end{array}
\end{equation}
where $P_{\beta=0}$ is the complex polarization corresponding to $F_c(\phi)$ when $\beta=0$. 
Changing the variable from $\pi\lambda^2$ to $\lambda^2$, we obtain
\begin{equation}
P(\lambda^2) = 
P_{\beta=0}(\lambda^2+\beta) 
e^{2{\rm i} \phi_{\rm 0} (\lambda^2 + \beta)} 
\,, 
\end{equation}
so that
\begin{equation}
|P(\lambda^2)| = |P_{\beta=0}(\lambda^2+\beta)| \,. 
\end{equation}
This shows that the observed modulus of the polarized intensity of a medium with a given $\beta$ is simply a 
translation by $\beta$ in $\lambda^2$-space of what would be observed if $\beta$ were equal to zero. 

$|F_c(\phi)|$ is real-valued, by definition; if it is even in $\phi$, its inverse Fourier transform is also real-valued, and the observed polarization angle will be 

\begin{equation}
\psi(\lambda^2) = \alpha + \phi_{\rm 0} (\lambda^2 + \beta) 
\end{equation}  
as found in Sect.~\ref{sect231} and \ref{sect232} for the top-hat and the Gaussian cases. 

\begin{figure}
\includegraphics[width=0.48\textwidth]{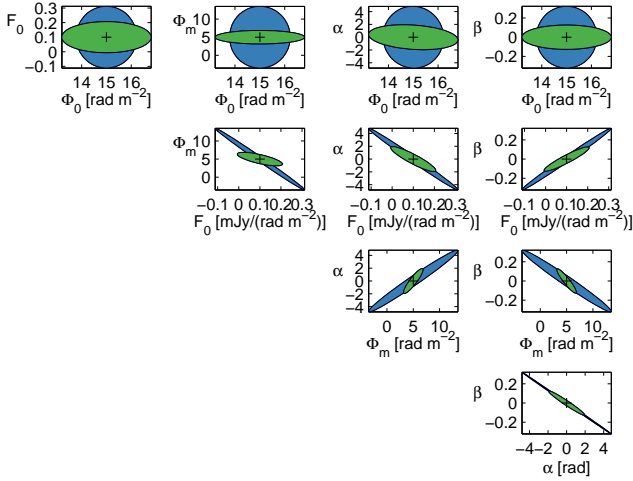}
\caption{
68.3 per cent confidence regions for a top-hat model of the Faraday dispersion function when 
$\beta = 0$. 
The ASKAP regions are in blue and the GMRT ones in green. 
}
\label{fig2}
\end{figure}

\begin{figure*}
\includegraphics[width=0.48\textwidth]{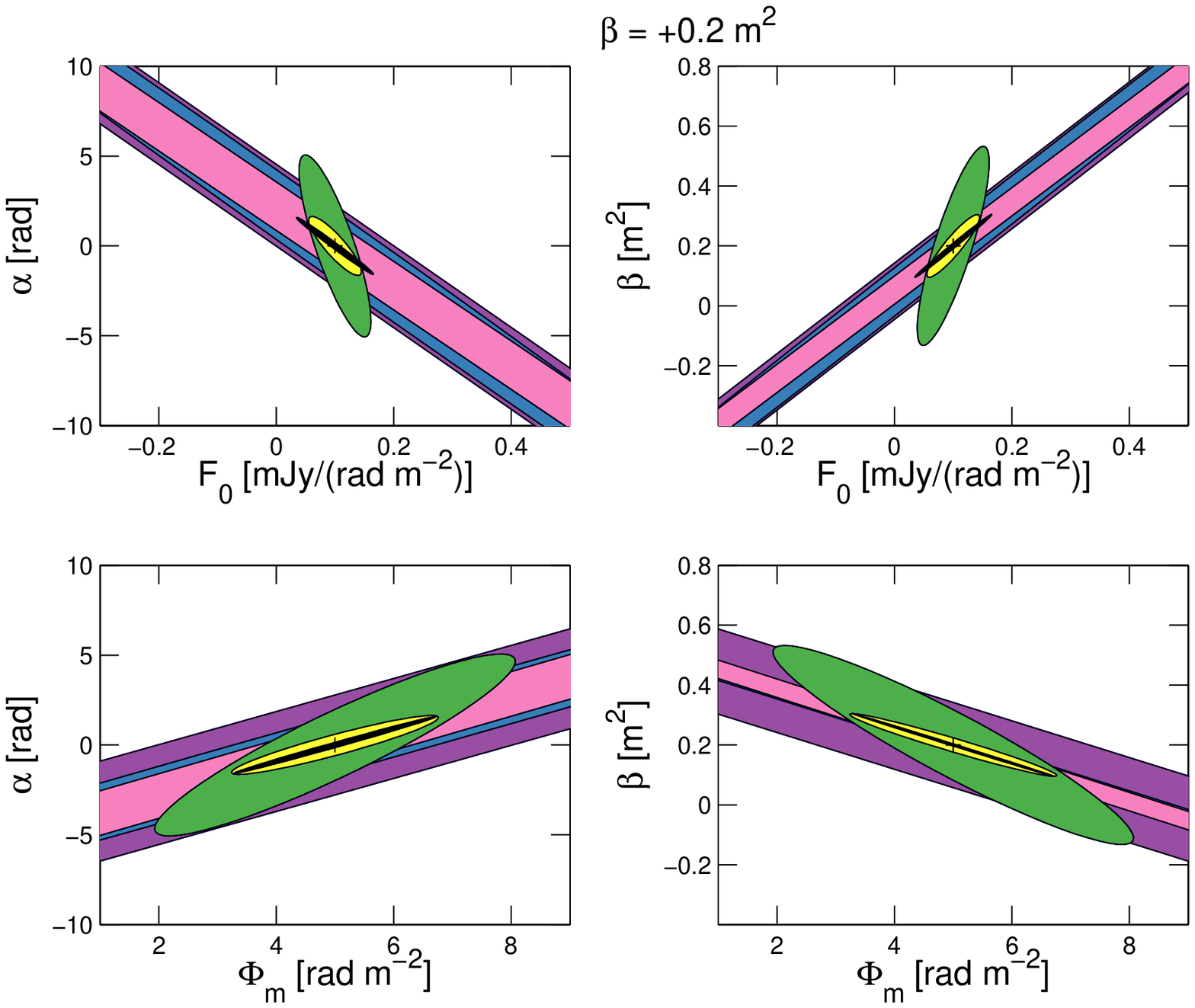}
\includegraphics[width=0.48\textwidth]{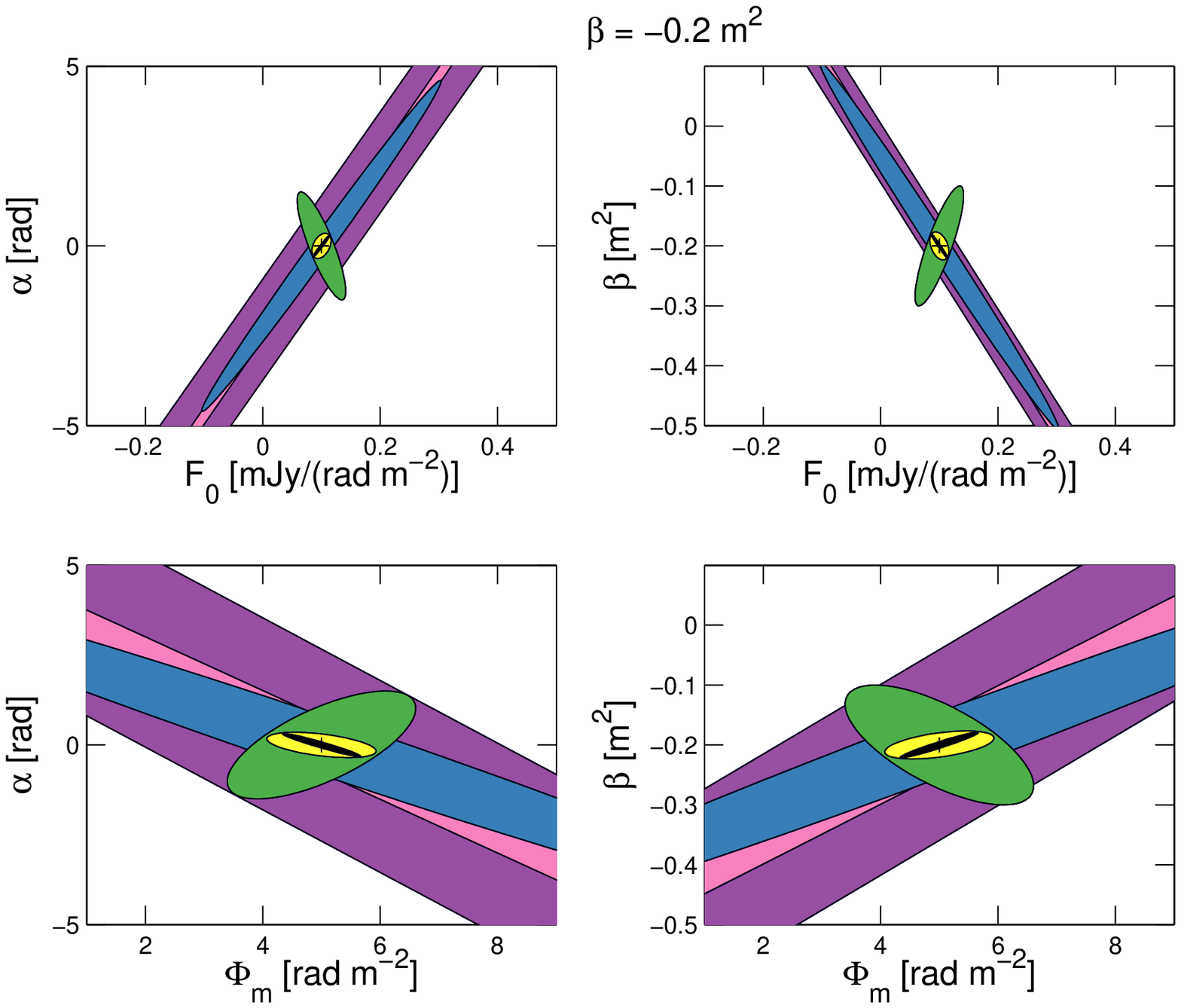}
\caption{
68.3 per cent confidence regions for a top-hat model of the Faraday dispersion function 
for some of the instruments listed in Table~\ref{tab1}. 
Left panel (four figures): $\beta = +0.2$~m$^2$; right panel (four figures): $\beta = -0.2$~m$^2$. 
The yellow ellipses show the confidence regions obtained by combining an ASKAP (blue ellipses) and a GMRT (green ellipses) data set. The SKA1-Survey confidence ellipses are shown in black, and the JVLA ones in purple (S-band) and pink (L-band). 
}
\label{fig3}
\end{figure*}

\subsection{Spectral dependence}

The expressions of $P(\lambda^2)$ used in the previous sections do not include any spectral dependence of the synchrotron emission. 
In general, the observed polarization can be written as the integral 
along the line of sight (los) and on a certain solid angle $\Omega_b$ on the sky of the 
intrinsic polarization modulated by the Faraday rotation, where the intrinsic 
polarization is a fraction of the total emissivity of the source: 
\begin{equation}
P(\lambda^2) = 
\int\displaylimits_{\rm los}\int\displaylimits_{\Omega_b} {\mathrm d}l {\mathrm d}\Omega  \, 
\epsilon(\bf{r},\lambda) \, p_0(\bf{r}) 
e^{2 {\rm i} (\psi_0({\bf r}) + \phi({\bf r})\lambda^2)}
\label{eqPwlambda} 
\end{equation}
where 
the emissivity in the direction ${\bf r}$, 
$\epsilon(r,\lambda)$, may depend on $\lambda$, and 
$p_0({\bf r})$ is the intrinsic degree of polarization. 

If the emissivity can be expressed as the product of two functions where one of them contains the spectral dependence 
(e.g., $\epsilon(r,\lambda) =  \hat{\epsilon}(r)\, s(\lambda)$), 
then $s(\lambda)$ can be taken out of the integral and 
Eq.~(\ref{eqPwlambda}) is invertible, following the same formalism at in Eqs~(\ref{eqpol}) and (\ref{eqF}) 
(for a discussion, see Sect.~3 of \citealt{2005A&A...441.1217B}). 

In the calculation of the confidence intervals that can be obtained from observations in various wavebands, we modelled 
the spectral dependence as a power law, 
\begin{equation}
s(\lambda) = \left(\frac{\lambda}{\lambda_0}\right)^{-\alpha_{\rm nth}} \, , 
\label{eqspecdep}
\end{equation}
with the normalisation wavelength $\lambda_0 = 0.2$~m and a fixed non-thermal spectral index, 
$\alpha_{\rm nth} = -1$. 
The values of the polarized intensity calculated using Eq.~(\ref{eqpol}) were multiplied by $s(\lambda)$.  
This effectively increases the flux densities measured at $\lambda > \lambda_0$ compared to the variations shown in Fig.~\ref{fig1}, where no spectral dependence was included.

\section{Results}

We use a fiducial top-hat model of the Faraday dispersion function 
with the following parameters: 
$\phi_{\rm 0}$ = 15 rad~m$^{-2}$, 
$\phi_{\rm m} = 5$ rad~m$^{-2}$,
$F_0$ = 0.1 mJy/(rad~m$^{-2}$), 
$\alpha = 0$~rad, and three values of 
$\beta$, 0 and $\pm0.2$~m$^2$. 
The total intrinsic polarized flux density (integrated over all Faraday depths) is thus $2\phi_{\rm m} F_0 = 1$~mJy. Because of depolarization, the signal expected at a given frequency will be weaker 
(see Figure~\ref{fig1}) but should be detectable within a reasonable amount of observing time  
by current and future instruments (see Table~\ref{tab1}). 
For comparison, we also use a Gaussian model of $|F(\phi)|$ with the same 
total flux. 
The dispersion of the Gaussian is $\sigma_\phi = \phi_{\rm m} = 5$~rad~m$^{-2}$ and the peak flux density per unit of Faraday depth $2/\sqrt{2\pi}F_0 \simeq 0.08$~mJy/(rad~m$^{-2}$). 
Note that $\sigma_\phi$ characterises the Gaussian profile of the Faraday dispersion function for a model with a regular field, not to be confused with 
the dispersion in rotation measure (RM) caused by possible RM fluctuations across an observing beam (usually denoted $\sigma_{\rm RM}$). 
In Sect.~\ref{sectFaradayDispersion} we discuss the additional depolarization effect by Faraday dispersion produced by a random field. 
The width of the Faraday structure (that is, the total Faraday depth, sometimes denoted $\mathcal{R}$, see Sect~\ref{sectFaradayDispersion}) 
in our fiducial top-hat model is $2\phi_{\rm m} = 10$~rad~m$^{-2}$.  
In many astrophysical cases this quantity can be larger (up to $\sim 80$ rad~m$^{-2}$ in spiral arms, e.g. \citealt{2011MNRAS.418.2336A}). 
A larger total Faraday depth translates into a narrower ``main peak" of the $P(\lambda^2)$ distribution and weaker emission at long wavelengths. 

\subsection{Differential Faraday rotation versus magnetic field helicity}

Fig.~\ref{fig1} shows the variation of the polarized intensity with $\lambda^2$ for a top-hat (left column) and a Gaussian (right column) Faraday dispersion 
function. In the former case, the solid line ($\beta = 0$) is the well-known sinc function produced by a uniform slab, which is more usually shown using the linear horizontal axis used in the bottom left panel. 
Note that, for clarity, the graphs in Fig.~\ref{fig1} do not include any spectral dependence of the intrinsic polarization. 
Figures~\ref{fig2} and \ref{fig3} (the confidence regions of the parameters) do, 
on the other hand, include a spectral dependence of the form given by Eq.~(\ref{eqspecdep}). 

In the Gaussian case, the polarized intensity decreases monotonically and 
no emission is produced in the longer wavebands at which the GMRT and LOFAR operate (Fig.~\ref{fig1}).
In the bottom right panel we show the variation of the $Q$ and $U$ Stokes parameters (in brown and blue) for the Gaussian $F(\phi)$, which are the direct observables. They oscillate with a 90$^\circ$ phase shift with respect to each other.

In the rest of the paper we focus on the top-hat model because it gives stronger
 emission in longer wavebands for the parameters selected here and it includes the standard case for Faraday depolarization calculations of a uniform slab 
as a special case. 
It is most interesting to compare the variations of the polarized intensity for 
a positive and a negative $\beta$.
When $\beta >0$, the intrinsic helicity of the magnetic field and the Faraday rotation 
act in the same direction. This results in an increased depolarization at short wavelengths. 
Even for $\lambda = 0$, where Faraday depolarization is absent, 
the emission is significantly depolarized compared to the case of a constant magnetic 
field orientation ($\beta = 0$). When $\beta < 0$, Faraday rotation counteracts the intrinsic rotation of the field, which means that the polarized emission peaks at a wavelength different from
 zero (dashed lines), where $\lambda^2 = -\beta$ (eqs.~(\ref{eqPtophat}) and (\ref{eqPgaussian})). This effect is similar to the ``anomalous depolarization'' discussed by \citet[Section 9]{1998MNRAS.299..189S}.

Fig.~\ref{fig2} shows the 68.3\% confidence regions of the five different parameters obtained from the Fisher analysis for $\beta = 0$. 
The colour code is the same as in Fig.~\ref{fig1}, with the ASKAP confidence regions in blue and the GMRT ones in green.  
The plots in the first row show that the central Faraday depth $\phi_{\rm 0}$ is mostly uncorrelated with the other parameters. 
On the other hand, $\phi_{\rm m}$, which describes the extent of the Faraday component in $\phi$-space, 
is strongly correlated with the parameters related to the intrinsic polarization angle ($\alpha$ and $\beta$) 
and at short wavelengths (in the ASKAP band) with the normalization of $F(\phi)$ ($F_0$). 
This is because the crucial effect is Faraday {\it differential} rotation across the Faraday component and not 
the magnitude of the central Faraday depth. An increase in $\phi_{\rm m}$ means stronger depolarization due to differential rotation 
which must be counteracted by a more negative $\beta$ in order to produce a similar fit to the data. Vice versa, a lower $\phi_{\rm m}$ means weaker depolarization and $\beta$ needs to become positive to increase the depolarization. In other words, $\phi_{\rm m}$ and $\beta$ are anti-correlated around $\beta = 0$. 
Most importantly, the strength of the correlation between pairs of parameters including $\alpha$ or $\beta$ 
varies with the selected waveband. This is what makes it possible to break parameter degeneracies by combining 
short-wavelength (like ASKAP) and long wavelength (like GMRT) data sets in order to better constrain the derived parameter values, as we discuss next. 

\begin{figure}
\includegraphics[width=0.48\textwidth]{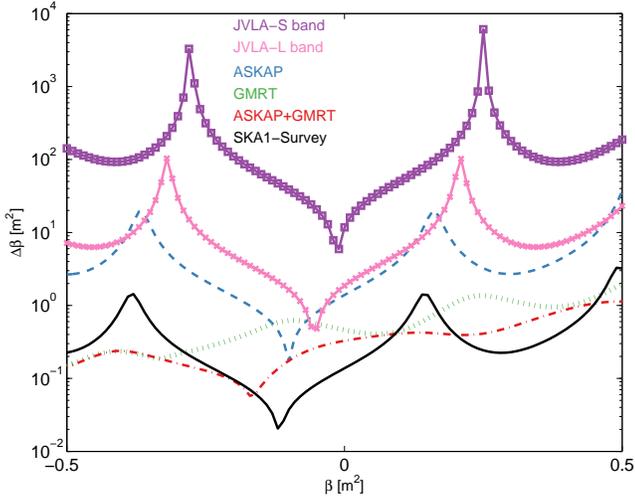}
\caption{Precision that can be achieved on the $\beta$ parameter as a function of $\beta$ 
for observations with some of the instruments listed in Table~\ref{tab1}. 
The combination of ASKAP and GMRT (shown here in dotted-dashed red) provides the smallest dependence of
the uncertainty on $\beta$, 
comparable to what can be achieved with SKA1-Survey. 
Note, however, the difference in integration time (a total of 3~h for the joint ASKAP/GMRT 
observations 
versus 10~min for SKA1-Survey).
}
\label{fig4}
\end{figure}

Fig.~\ref{fig3} shows the one-sigma confidence ellipses for 
$\alpha$, $\beta$, $F_0$ and $\phi_{\rm m}$ 
for $\beta = +0.2$~m$^2$ (four left panels) 
and $\beta = -0.2$~m$^2$ (four right panels). 
As expected, the orientation of the ellipses is similar for the short-wavelength instruments JVLA (S-band in purple and L-band in pink), ASKAP (blue) and SKA1-Survey (black). 
On the other hand, the long-wavelength GMRT data set (green) produces a different correlation between parameters; in some cases the confidence ellipses at short and long wavelengths are almost orthogonal, making it possible to reduce the confidence intervals on the derived parameter values considerably by using both wavelength ranges together (such as ASKAP and GMRT, shown in yellow). 
This is further illustrated in Fig.~\ref{fig4} where the precision that can be achieved in $\beta$ for different instruments is shown. A combination of ASKAP and GMRT observations makes it possible to reach an uncertainty $\Delta\beta < 0.25$~m$^2$ if $\beta < 0$, for the integration times shown in Table~\ref{tab1} and the set of model parameters used. If $\beta >0$, all signals are weaker because of increased depolarization and the precision on $\beta$ (and all other model parameters) is lower.

\subsection{Faraday dispersion}  
\label{sectFaradayDispersion}

\begin{figure}
\includegraphics[width=0.48\textwidth]{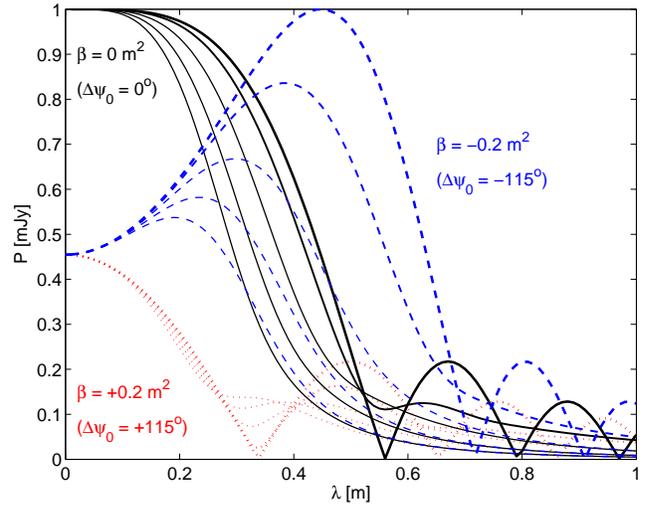}
\caption{Variation of the polarized intensity with wavelength in the case of a uniform slab   
of Faraday depth $\mathcal{R} = 10$~rad~m$^{-2}$ and Faraday dispersion characterised by $\sigma_{\rm RM} = 0, 2.5, 5, 7.5$ and 10~rad~m$^{-2}$ (curves from top to bottom) for $\beta = 0$ (no helicity; black solid lines), 
$\beta = +0.2$~m$^2$ (red dotted lines) 
and $\beta = -0.2$~m$^2$ (blue dashed lines).
The two latter cases correspond to a gradual change of the perpendicular component of the magnetic field by about $115^\circ$ along the line of sight across the slab. 
The presence of a random field attenuates the variations and shifts the peak of the polarized emission towards shorter wavelengths. 
} 
\label{figFdispersion}
\end{figure}

In this section we discuss the effect of depolarization by Faraday dispersion, due to  
possible random fluctuations of the magnetic field inside the synchrotron-emitting source. 
This effect was first discussed by \cite{1998MNRAS.299..189S} (Sect.~9) 
but they considered only the case of twisted fields where the rotation due to the helicity is counteracted by the Faraday rotation ($\beta < 0$ in our notation). 
As detailed in Appendix~\ref{app1}, the observed complex polarized emission 
is 
\begin{equation}
\langle P(\lambda^2) \rangle = 
e^{2{\rm i} \Delta\psi_0}
\frac{1-e^{-S_H}}{S_H} 
\end{equation}
where 
\begin{equation}
S_H = -2 {\rm i} \left( 
\mathcal{R} \lambda^2 - \Delta\psi_0 \right)
+ 2 \sigma_{\rm RM}^2 \lambda^4 \, .
\end{equation}
$\Delta\psi_0 =  k_H L$ is the total rotation due to the helicity of the intrinsic polarization angle $\psi_0$ across the thickness $L$ of the slab, $\mathcal{R}$ is the total Faraday depth of the source 
and $\sigma_{\rm RM}$ is the dispersion of the total Faraday depth. 
 In the previous sections we had $\mathcal{R} = 2\phi_{\rm m} = 10$~rad~m$^{-2}$ 
and $\Delta\psi_0 = 2\phi_{\rm m} \beta = 0$ and $\pm 2$~rad $\simeq \pm 115^\circ$. 
Figure~\ref{figFdispersion} shows the variation of the polarized intensity with wavelength for the three cases considered before (a non-helical magnetic field ($\beta = 0$) and helical magnetic fields with $\beta = \pm 0.2$~m$^2$) 
to which the effect of a random magnetic field component was added. 
The random fluctuations are described by values of $\sigma_{\rm RM}$ increasing from 0 (top, thicker curves) to 10~rad~m$^{-2}$ by step of 2.5~rad~m$^{-2}$. 
The Faraday dispersion caused by the random fluctuations 
attenuates the variations and, for negative values of $\beta$ where the intrinsic helicity acts in the opposite direction of the Faraday rotation, shifts the 
peak of the polarized intensity towards shorter wavelengths. 

\section{Summary}

We examined how variations of the intrinsic polarization angle  $\psi_{0}$ with the Faraday depth $\phi$ within a source affect the observable quantities. Using simple models for the Faraday dispersion $F(\phi)$ and $\psi_{0}(\phi)$, along with the current and planned properties of the main radio interferometers, we show
how degeneracies among the parameters describing the magneto-ionic medium can be minimised by combining observations in different wavebands. In particular we have shown that it may be possible to recover the sign and the magnitude 
of $\beta$, 
a parameter that we have defined and that is related to the relative effect of the helicity of the transverse magnetic field and the Faraday rotation due to the parallel component of the magnetic field. Since the direction of $B_{||}$ can be easily inferred from RM measurements, it should be possible to recover the sign (and, under some assumptions the magnitude) 
of the helicity of the magnetic field. 
However, the additional effect of Faraday dispersion by a random component of the magnetic field 
attenuates the variations of the polarized emission as a function of wavelength and may shift the peak 
of polarized emission towards shorter wavelengths if $\beta$ is negative. Faraday depolarization effects 
(both by differential rotation and by dispersion) will have to be included in the modelling of real data in order to recover 
information on the helicity of the magnetic field. 
This approach is complementary to statistical studies of the correlation between the degree of polarization and the rotation measures of cosmic sources which may also provide information on magnetic helicity (\citealt{2010JETPL..90..637V}, \citealt{2014arXiv1401.4102B}). 

Planned surveys of fixed sensitivity will be biased towards radio sources with negative $\beta$, because of the depolarization produced when $\beta > 0$. Detection of $\beta > 0$ will be more difficult both through $Q(\lambda^2)$ and $U(\lambda^2)$ model-fitting and RM synthesis because most of the signal is shifted toward negative 
$\lambda^2$. \cite{2014arXiv1401.4102B} 
show that restricting the integral in Eq.~(\ref{eqF}) to the 
positive $\lambda^2$ yields an erroneous reconstruction of $F(\phi)$ when $\beta >0$.  

In this work we used a first-order parametrisation of the variation of intrinsic polarization angle with Faraday depth. Higher-order representations could be used (or e.g. Chebyshev polynomials) if the data are of sufficient quality. Including a second-order term implies a convolution with an imaginary Gaussian 
and a significantly more complicated expression for $P(\lambda^2)$.

\section*{Acknowledgments}
We thank Oliver Gressel for organizing the stimulating Nordita workshop `Galactic Magnetism in the Era of LOFAR and SKA' held in Stockholm in 2013 September . 
We also thank Axel Brandenburg and Rodion Stepanov for interesting discussions on the topics discussed in this paper and for sharing their related results with us. 
We are grateful to Rainer Beck, John~H. Black, Jamie Farnes, George Heald, Anvar Shukurov and to the referee for constructive comments. 
AF is grateful to the Leverhulme Trust for financial support under grant RPG-097.

\appendix 
\section{Faraday dispersion around a mean helical field}  \label{app1}

In general, Eq.~(\ref{eqpol}) can be rewritten as an integral along the line of sight of the intrinsic complex polarization modulated by the Faraday rotation: 
\begin{equation} 
P(\lambda^2) = \int_{\rm source}^{\rm observer} {\mathrm d}z 
F(z) \, e^{2{\rm i}\phi(z)\lambda^2}\, .
\end{equation}

For a slab-like source of intrinsic polarized emission $|F(z)| = 1/L$ between $0\leq z \leq L$  
and intrinsic polarization angle $\psi_0(z) = k_H z$, it becomes

\begin{equation} 
P(\lambda^2) = \frac{1}{L} \int_0^L {\mathrm d}z \, 
e^{2{\rm i}(k_H z + \phi(z)\lambda^2) } \, .
\end{equation}

Our system of coordinates is chosen so that the origin is at the far end of the source and the observer is placed at $+\infty$ (note that 
\citealt{1966MNRAS.133...67B} 
placed the observer at the origin, 
where \citealt{1998MNRAS.299..189S} 
used the symmetry plane of the slab as the origin of the reference frame, 
so that the source would extend from $-L/2$ to $+L/2$. 
The choice of the origin of the reference system is of course arbitrary, but it introduces a phase in the final expression of the observed complex polarization). 

We consider now the additional effect of Faraday dispersion produced by a random component of the magnetic field. 
The observed complex polarized intensity is 
\begin{equation} 
\langle P(\lambda^2)\rangle  = 
\frac{1}{L} \int_0^L {\mathrm d}z \, 
e^{2{\rm i} k_H z} \, 
\langle e^{2{\rm i}  \phi(z)\lambda^2}\rangle \, , 
\label{eqkH}
\end{equation}
where $\langle \rangle$ denotes the ensemble average of the quantity between brackets: 
\begin{equation} 
\langle  e^{2{\rm i}  \phi(z)\lambda^2} \rangle = 
  \int_{-\infty}^{+\infty} {\mathrm d}\phi \, 
p(\phi)\, e^{2{\rm i} \phi(z) \lambda^2 } \, , 
\label{eqAverageFaradayTerm}
\end{equation}
where $p(\phi)$ is the probability distribution function (PDF) of $\phi$. 

The total magnetic field is the superposition of a regular component, $\mathbf{B}$, and a random one, $\mathbf{b}$. The scale on which the random field varies is denoted $d$. The components of the magnetic field along the line of sight are denoted 
\begin{equation}
{\bf B_{{\rm tot}\parallel}} = {\bf B_{\parallel}} + {\bf b_{\parallel}}\, .
\end{equation}

Let us calculate $p(\phi)$. The Faraday depth at a given location $z$ along the line of sight is 
\begin{equation} 
\begin{array}{ccccc}
\phi(z) &= &0.81 n_e \int_z^L {\mathrm d}z \, 
(B_{\parallel} &+ &b_{\parallel}) \\ 
        &= &\underbrace{m (L-z)} 
&+ &\underbrace{0.81 n_e \int_z^L {\mathrm d}z \, b_{\parallel}} \\
	&  &{\rm regular} &    & {\rm random}
\end{array}
\end{equation}
where $m = 0.81 n_e B_{||} = \mathcal{R}/L$, where $\mathcal{R}$ is 
the total Faraday depth of the source. 
If $b_{||}$ is a Gaussian variate, then $\phi(z)$ is also a Gaussian  
variate\footnote{Note that \cite{1966MNRAS.133...67B} 
had a factor $d/2$ instead of $d$ in his expression of the variance, which resulted in a factor of 2 in the ``random" term of $S$ (see Eq.~(\ref{eqSH})), as also noted by \cite{1998MNRAS.299..189S}.}, $p_G(\phi)$,  
of mean $m(L-z)$
and variance $(0.81n_e b_{\parallel}d)^2 \left( \frac{L-z}{d} \right) = v^2 (L-z)$, 
where $v^2 = (0.81 n_e b_{\parallel})^2 d$. Equation~(\ref{eqAverageFaradayTerm}) becomes 
\begin{equation} 
\begin{array}{lrcc}
\langle e^{2{\rm i}  \phi(z)\lambda^2}\rangle  = 
\int_{-\infty}^{+\infty} &{\mathrm d}\phi \, 
e^{2{\rm i} \phi\lambda^2} \\ 
&p_G\big( \phi;  
 &\underbrace{\frac{\mathcal{R}}{L}(L-z)};  
 &\underbrace{v^2(L-z)} 
                     \big) \, .\\ 
& 
&{\rm mean} 
&{\rm variance} 
\end{array}
\end{equation}
This can be expressed as the inverse Fourier transform of the Gaussian PDF, which can be rewritten in a more convenient manner using the properties of the Fourier transform: 
\begin{equation} 
\begin{array}{lcl}
\langle e^{2\pi{\rm i}  \phi(z)\lambda^2}\rangle 
&= 
&{\rm FT}^{-1}\big\{ 
p_G\big(\phi; 0; v^2(L-z)\big) 
* \delta\big( \phi - \frac{\mathcal{R}}{L}(L-z) \big)
\big\} \\
  \\ 
&= 
&\exp[ - \frac{(2\pi\lambda^2)^2 v^2 (L-z)}{2}]  
\cdot e^{2\pi{\rm i} \frac{\mathcal{R}}{L}(L-z)\lambda^2} 
\end{array}
\end{equation}
Replacing $\pi\lambda^2$ by $\lambda^2$, and simplifying: 
\begin{equation} 
\begin{array}{ll}
\langle e^{2{\rm i}  \phi(z)\lambda^2}\rangle 
&= \exp[ 
-\frac{S}{L}(L-z)] \, , 
\end{array}
\end{equation}
where 
\begin{equation} 
\begin{array}{llccc}
S &= 
& \underbrace{- 2{\rm i} \mathcal{R} \lambda^2 } 
&+ &\underbrace{2 v^2 L \lambda^4} \, . \\
& 
&{\rm regular} & & {\rm random} 
\end{array}
\end{equation}
Finally Eq.~(\ref{eqkH}) becomes  
\begin{equation} 
\begin{array}{lcl}
\langle P(\lambda^2) \rangle &= &\frac{1}{L} \int_0^L {\mathrm d}z\, 
e^{2 {\rm i} k_H z} \, \exp[-\frac{S}{L}(L-z)] \\
\\ 
&= &e^{2{\rm i} \Delta\psi_0} \left( \frac{1-e^{-S_H}}{S_H} \right)\, ,
\label{eqPdisp}
\end{array}
\end{equation}
where $\Delta\psi_0 = k_H L$ is the total rotation of the polarization angle across the slab and 
\begin{equation}
S_H = -2 {\rm i}\, (\mathcal{R}\lambda^2 - \Delta\psi_0) + 2 v^2 L \lambda^4\, .
\label{eqSH}
\end{equation}
Equation~(\ref{eqPdisp}) is a generalization 
of the expressions provided by \cite{1966MNRAS.133...67B} 
and \cite{1998MNRAS.299..189S} 
to the case of Faraday dispersion around an helical field.  
$v^2 L$ is the variance of the total Faraday depth of the source and is usually denoted $\sigma_{\rm RM}^2$. 


\end{document}